%% file: jhepMay2016.tex
\documentclass[a4paper,11pt]{article}
\pdfoutput=1 

\usepackage{jheppub} 

\usepackage[T1]{fontenc} 

\usepackage{graphicx}

\title{\boldmath The Emergence of Superconducting Systems in Anti-de Sitter Space}

\author[a]{W.M. Wu}
\author[b]{M.P. Pierpoint,}
\author[c,1]{D.M. Forrester\note{Corresponding author.}}
\author[a]{and F.V. Kusmartsev}

\affiliation[a]{Department of Physics, Loughborough University, Loughborough LE11 3TU,
United Kingdom}
\affiliation[b]{Institute of Physics and Chemistry, Ogarev Mordovia State University, 
Saransk, 430005, Russia}
\affiliation[c]{Department of Chemical Engineering, Loughborough University, Loughborough LE11 3TU,
United Kingdom}

\emailAdd{W.M.Wu2@lboro.ac.uk}
\emailAdd{M.Pierpoint@lboro.ac.uk}
\emailAdd{D.M.Forrester@lboro.ac.uk}
\emailAdd{F.Kusmartsev@lboro.ac.uk}

\abstract{In this article, we investigate the mathematical relationship between a (3+1) dimensional gravity model inside Anti-de Sitter space $\rm AdS_4$, and a (2+1) dimensional superconducting system on the asymptotically flat boundary of $\rm AdS_4$ (in the absence of gravity). We consider a simple case of the Type II superconducting model (in terms of Ginzburg-Landau theory) with an external perpendicular magnetic field ${\bf H}$. An interaction potential $V(r,\psi) = \alpha(T)|\psi|^2/r^2+\chi|\psi|^2/L^2+\beta|\psi|^4/(2 r^k )$ is introduced within the Lagrangian system. This provides more flexibility within the model, when the superconducting system is close to the transition temperature $T_c$. Overall, our result demonstrates that the two Ginzburg-Landau differential equations can be directly deduced from Einstein's theory of general relativity.}

\begin{document} 
\maketitle
\flushbottom

\section{Introduction}
\label{sec:intro}
The lattice structures of superconducting cuprates or pnicides are characterised by complex competing electronic and magnetic phases that emerge in association with fractal structures that develop from the nano level, and propagate up to many micrometres in size \cite{Bianconi2010}. To create a high temperature superconductor one needs to dope a parent compound such as $La_2CuO_4$ with, for example, oxygen interstitials or strontium. The doping that creates the highest critical temperature is called optimal doping. At this optimal level a single $T_c$ value marks the transition to a superconducting phase. However, careful annealing to avoid the escape of interstitial oxygen produces a mixed state that can even have two critical temperatures \cite{Bianconi2010}. This is caused by the self-organisation of the oxygen into different patterns, such as stripes \cite{Kusmartsev2000}, or the formation of dipolar resonance plaquettes \cite{Kusmartsev2015}. Thus, a $1$ or $2D$ ordering of electronic density in high temperature superconductors may dictate the properties of the phase diagram and in particular the superconducting phase. At the optimal doping a continuous phase change - with a quantum critical state - may be realised at the transition point \cite{Marel2003}. In these high temperature superconductors electron pairs form, but the mechanisms are quite different from conventional superconductors (that generally have much lower critical temperatures). There is a strong coupling mechanism (that is not due to phonons) involved that renders well-known theories that make use of the BCS model (after Bardeen Cooper and Schrieffer) unable to describe the physical properties. Conventional superconductivity typically involves pairs of electrons that are separated over distances larger than the lattice spacing, leading to a relatively weak binding. Thus, new methods of analysis are required in condensed matter physics that can lead to greater understanding of the complex issue of strongly coupled systems. 

Here, we make use of techniques borrowed from cosmology that are valid for describing a superconductor when its temperature is equivalent to that of a corresponding black hole \cite{Horowitz2011}. The fractal structures found in the high $T_c$ superconductors contain information about the origin and history of the sample. Likewise, a black hole's information is contained in threads or ``hair'' at its event horizon that grow from its time of formation, and its later development \cite{Hawking2016}. To use the cosmological models we need to mathematically create a black hole that has hair below $T_c$. The emergence of the superconducting phase corresponds to a black hole formed in Anti de Sitter ($AdS$) space~\cite{Hawking,Barbon,Hartle} with hair \cite{Erdmenger2013}. A quantum critical point ($QCP$) is suspected to lie within the superconducting phase and quantum fluctuations are thought to extend its presence to temperatures well above absolute zero. Near $T_c$ the quantum fluctuations should be detectable throughout the superconducting condensate, with analogy to a black hole with the same quantum hair (i.e. entropy, information, temperature) \cite{DvaliGomez2014}.  Recent work in the iron pnictide superconductors \cite{Hashimoto2012} has found a $QCP$ where the London penetration depth increases as a consequence of quantum fluctuations.  A further signature of a $QCP$ is superconductivity and magnetism coexisting as a consequence of doping. A possible material for demonstration of this phenomena is the new material with anomalous magnetoresistance, $Li Ti_2 O_4$ \cite{Jin2015}. The work we develop herein may be useful in further developing the understanding of the physics of these novel materials. Currently investigations into whether the holographic principle (mapping the system of interest to one dimension higher) is a fundamental aspect of the character of the Universe are being carried out at the Fermilab Holometer (holographic interferometer) facility. The objective is to find evidence of quantum fluctuations at extremely low temperatures and to establish space-time as a quantum system. Also of associated interest are recent developments in measuring quantum fluctuations in an opto-mechanical system using superconducting Josephson junctions in parallel with an $LC$ oscillator operating at a single quantum level \cite{Lecocq2015}. This set-up is analogous to the Fermilab Holometer in many respects, within the confines of a superconducting system. 

The high $T_c$ superconductors are layered and can be described by ($2+1$) dimensional models. Using the $AdS$ gravitational model, the properties in the vicinity of ($2+1$)  quantum critical points may be investigated by finding a ($3+1$)-dimensional gravitational dual of the ($2+1$) dimensional system below $T_c$ . The $AdS$ space is becoming an increasingly valuable tool in different branches of physics - including cosmology, string theories~\cite{Maldacena1,Merali,Bousso3,Wilczek,Yin,Paul,Hartle2}, condensed matter physics~\cite{Gubser,Horowitz,Hartnoll,Garcia}, and more recently, within the holographic principle~\cite{Maldacena1,Klebanov,Bagchi,Bousso1,Bousso2}. AdS space has a negative curvature, conveniently offering resolution to the problem of the thermodynamically unstable Schwarzschild black-hole, as it possesses negative heat capacity \cite{Bousso1,Bousso2}. Thus, we provide a new methodology that demonstrates that the application of an $AdS$ space within Einstein's theory of relativity can lead to the emergence of a superconducting system (i.e., Ginzburg-Landau theory \cite{Ginzburg,Abrikosov}), which is located on the AdS infinite boundary~\cite{Gubser,Horowitz,Hartnoll,Garcia}.

\section{Gravity Model}
The line element of AdS$_4$ is given by~\cite{{Garcia}}  
\begin{equation}
ds^2 = g_{\mu \nu} d x^{\mu} dx^{\nu},
\end{equation}
with the system defined by Poincare coordinates $x^{\mu}=\{t,r,x,y\}$ \cite{Braga}.
The metric $g_{\mu\nu}$ is chosen as 
\begin{displaymath}
g_{\mu \nu} =
 \begin{pmatrix}
-s(r)& 0 & 0 & 0 \\
0 & {1}/{s(r)} & 0 & 0 \\
0 & 0  & r^2 & 0  \\
0 & 0 & 0 & r^2
 \end{pmatrix}
\end{displaymath} 
with the function~\cite{Horowitz,Hartnoll,Garcia}
\begin{equation}
s(r) = \frac{r^2}{L^2}-\frac{r_0^3}{L^2 r} .
\end{equation}
Here $r_{0}$ represents the horizon radius of the AdS black-hole, and is directly related to its mass. To be consistent with dimensionality, the characteristic AdS length scale $L=\sqrt{-3/\Lambda}$ has been re-parametrized in terms of the cosmological constant $\Lambda$ \cite{Horowitz,Garcia}. We also note that the function $s(r)$ is not chosen arbitrarily, but rather as an explicit solution to the Einstein equations \cite{Horowitz,Hartnoll,Garcia}. Both the AdS length scale $L$ and horizon radius $r_0$ determine the black-hole temperature $T = \frac{3r_0}{4\pi L^2}$, as mentioned by Hawking \cite{Hawking,Garcia}. 
\begin{figure}
\begin{center}
\includegraphics[width=8.5cm]{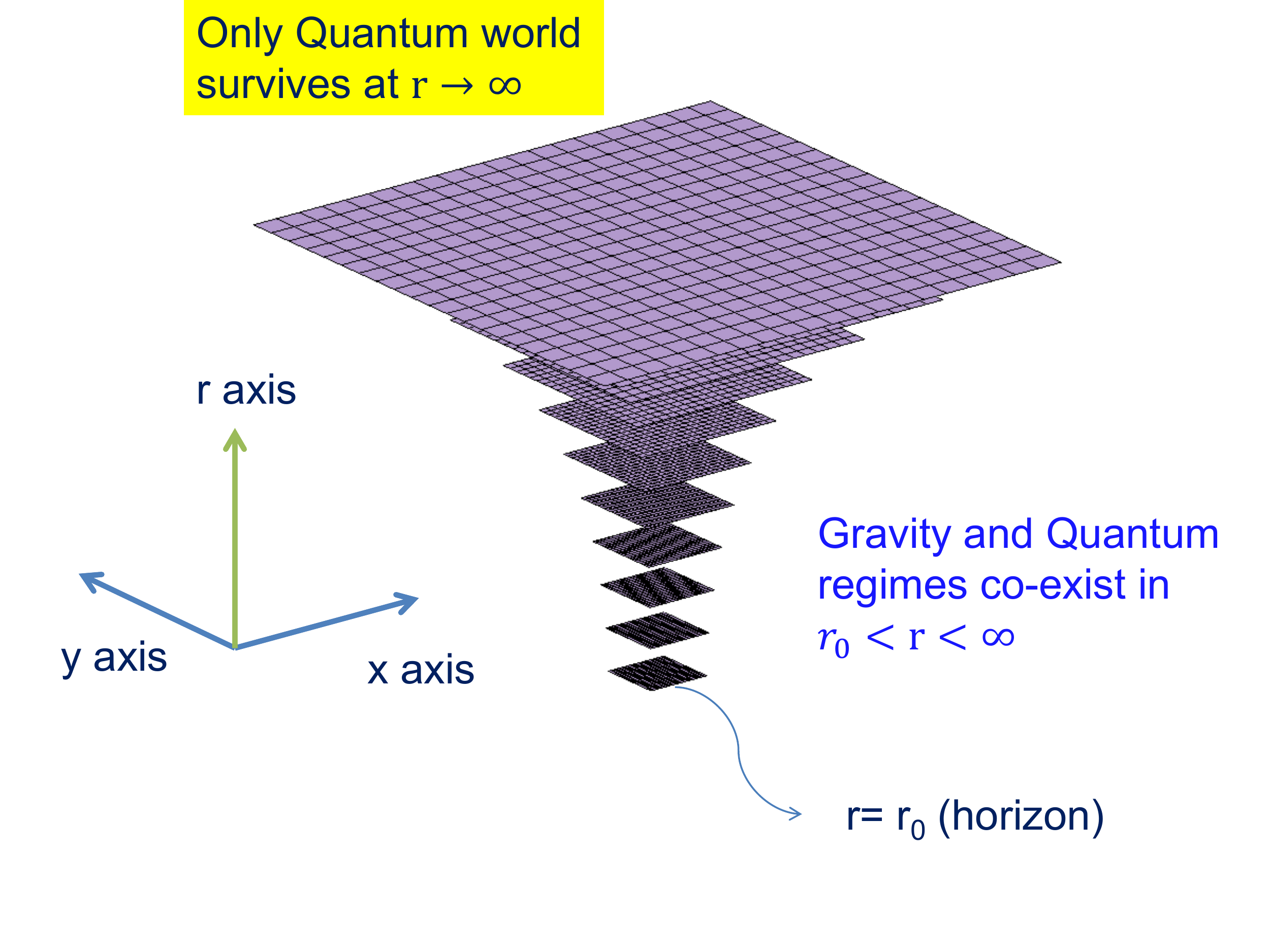}
\caption{\label{geometry} The spatial geometry in Anti-de Sitter (AdS) space with coordinates $(r,x,y)$ is shown. Here, $r$ represents the holographic axis, with only quantum phenomena surviving on the asymptotically flat (x-- y plane) at $r\rightarrow \infty$.}
\end{center}
\end{figure}

The scale of the x--y coordinate plane increases with the square of the holographic dimension $r$, namely $r^2(dx^2+dy^2)$.
Assuming a stationary system with negligible back-reaction, the spatial geometry of AdS can be realized as in Fig~\ref{geometry}. Back-reaction here refers to the curvature of space-time induced by small particles.

The required action for the gravity model is given by $S=\int\mathcal{L}\sqrt{|g|}\,{\rm d}^4x$, where the Lagrangian density $\mathcal{L}$ \cite{Montull,Herzog,Maeda} is given by
\begin{equation}
\mathcal{L} = \frac{1}{2\kappa}\left( \mathcal{R} + \frac{6}{L^2}\right) + \mathcal{L}_m ,
\end{equation}
and 
\begin{equation}
\mathcal{L}_m =  - \frac{1}{4 \mu_0}F_{\mu\nu} F^{\mu\nu} - \frac{\hbar^2}{2 m^*}|(\nabla - i\frac{q}{\hbar}{\bf A}) \psi|^2  - V(r,\psi).
\end{equation}
$\mathcal{L}_m$ is a Lagrangian density for matter fields, where $\mu_0$ is the permeability of free space, $\hbar$ is Planck's constant, $\bf A$ is magnetic vector potential, $q$ represents charge, and $m^*$ is the effective mass of a charge. The action $S$ is a functional of a (complex) scalar field $\psi$, and includes the Ricci scalar curvature $\mathcal{R}$, the determinant $g=\det(g_{\mu\nu})$, the gravitational coupling constant $\kappa = 8\pi G$, electromagnetic fields $F^{\mu\nu}$ and an interaction potential $V(\psi)$.
\begin{figure}
\begin{center}
\includegraphics[width=8.5cm]{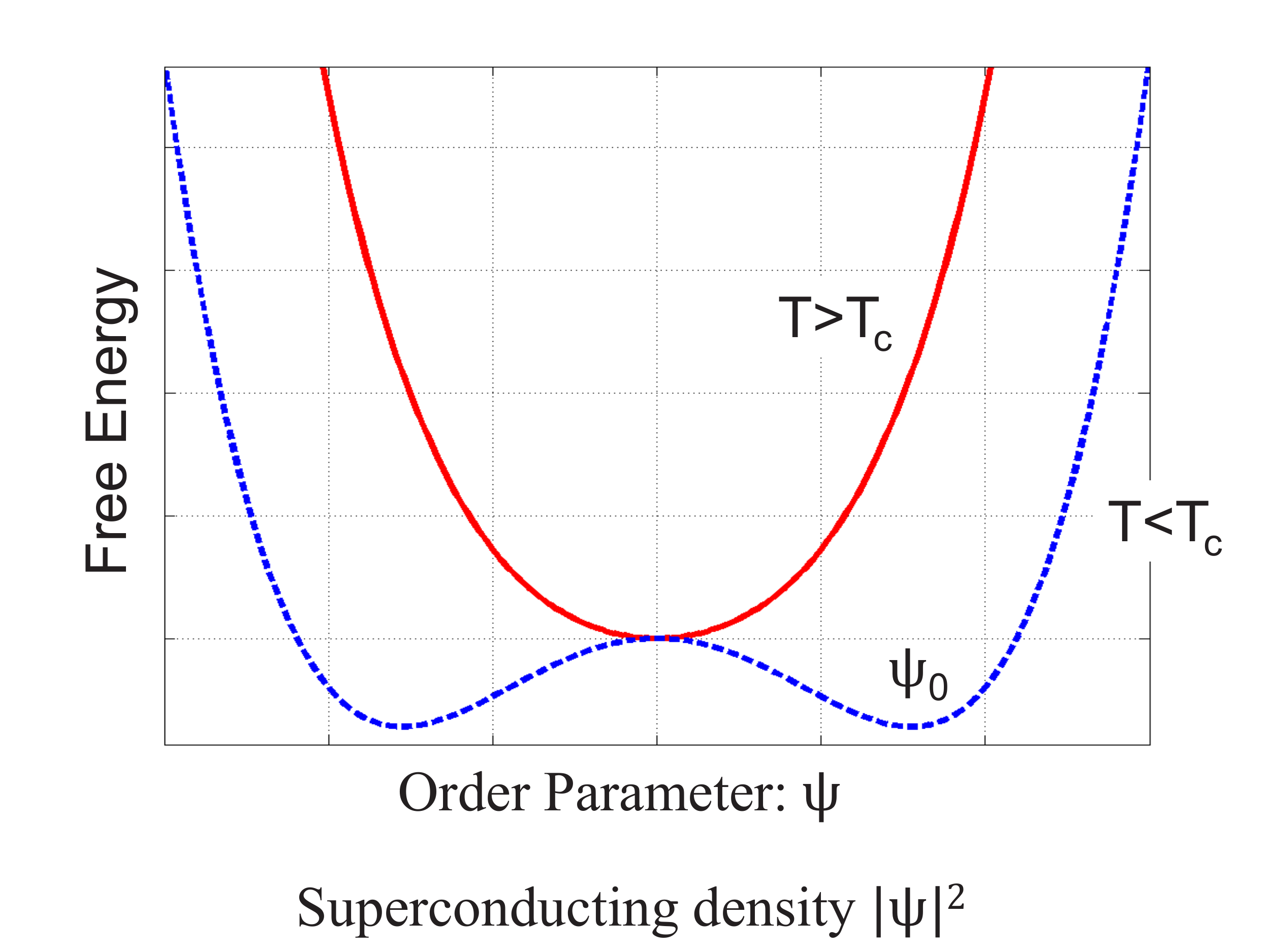}
\caption{\label{phase_tran} The free energy associated with a phase transition of superconducting density $|\psi|^2$ is shown. For temperature $T>T_c$, the minimum free energy is located at $\psi = 0$ - corresponding to the normal state ($|\psi|^2=0$). For $T<T_c$, a minimum free energy is now positioned at $\psi = \psi_0$ - representing the superconducting state ($|\psi|^2>0$) which refers to symmetry breaking.}
\end{center}
\end{figure}
\section{Superconducting system}
Our ultimate goal, is to establish a relationship between the (3+1) gravity model in the bulk, and the (2+1) Ginzburg-Landau theory upon the AdS boundary. We incorporate a flexible solution for the potential as follows
\begin{equation}
V(r,\psi) = \alpha(T)\frac{|\psi|^2}{r^2}+\chi\frac{|\psi|^2}{L^2}+\beta\frac{|\psi|^4}{2 r^k},
\end{equation}
where $k$ is an integer (dependent upon the choice of solution), $\chi$ and $\beta$ are constants, and $\alpha(T) \propto (T-T_c)$  is a temperature dependent parameter. It is clear that this parameter $\alpha(T)$ changes sign at some critical temperature $T_c$, and corresponds to the phase transition described by the Ginzburg-Landau theory~\cite{Tinkham,Gennes,Schweigert,James}. For $T<T_c$, this corresponds to a superconducting state; whereas for $T>T_c$, this implies a normal state. 
Fig.~\ref{phase_tran} shows the idea of the phase transition of a superconducting system, associated with symmetry breaking~\cite{James}. We also have to mention that our proposed model is not scale invariant due to the choice of potential $V(r,\psi)$.

Generally speaking, superconductors can be classified as either Type I or Type II. The Type I superconductors are famous for the Meissner effect \cite{Dorsey}. As $H$ exceeds a critical value $H_c$, a normal state emerges. For Type II superconductors, there are the two critical limits, $H_{c1}$ and $H_{c2}$. For a magnetic field less than $H_{c1}$ or greater than $H_{c2}$, the superconductor is either in the Meissner state or normal state respectively. For vortices nucleating in between the two critical limits, $H_{c1}<H<H_{c2}$, we call this a mixed state \cite{Abrikosov}. For a discussion of vortices in Type II superconductors and holography, see references \cite{Roychowdhury2014a,Roychowdhury2014b,Roychowdhury2015,Rogatko2015b}.

Generally, the BCS theory \cite{Gennes} describes the Type I superconductors, and likewise, the Ginzburg-Landau theory can explain Type II superconductors near the point of phase transition. It is due to this fact, that Ginzburg-Landau theory is capable of solving the strong coupling of two non-linear differential equations \cite{Tinkham}. These coupled equations can resolve the complex scalar field and magnetic potential $(\psi,{\bf A})$, respectively. In this study, we focus our attention upon the Type II superconductors for a two-dimensional geometry \cite{Chibotaru,Mel'nikov,Abrikosov}.

\section{Methodology and Analysis}
We will now go on to establish the link of a scalar field $\psi$ of the gravity model in the bulk, to a quantum wave function ($\psi = |\psi| e^{i\phi}$) of the Ginzburg-Landau model \cite{Abrikosov} at the boundary of AdS. We also assume the absence of back-reaction on the infinite boundary ($r \rightarrow \infty$) of AdS~\cite{Gubser} (back-reacting holographic superconductors were discussed in References \cite{Rogatko2014,Wysokinski2015,Nakonieczny2015,Rogatko2015a}). 

We consider a small 2-D superconducting system with an applied static magnetic field $H_{\perp}$ - acting perpendicular to the x--y coordinates at the AdS boundary. We also assume the superconducting state to be stationary, with no overall time dependence of the system. Moreover, we are able to choose the gauge to be $\partial_x A_x = 0$ and $\partial_y A_y=0$. As such, we can set the parameters $A_{t} = 0$, $A_r = 0$, with all time and holographic components vanishing also. Therefore, $H_{\perp} = \nabla_{(x,y)} \times {\bf A} = \partial_y A_x - \partial_x A_y$. Here, $\nabla_{(x,y)}$ is defined as acting upon the x--y coordinate only, with gauge field ${\bf A}=\left\{A_{x},A_{y}\right\}$. The coupled parameters are all functions of coordinates $r$, $x$ and $y$; these are the magnetic potential $A_x(r,x,y)$, $A_y(r,x,y)$ and scalar field $\psi(r,x,y)$.

\subsection{Coupling Differential Equations}
Following from the Euler-Lagrange equations \cite{Lewis}, 
\begin{align*}
\frac{\partial (\mathcal{L}\,\sqrt{|g|})}{\partial \psi^*} - \frac{d}{d x^{\mu}}\frac{\partial (\mathcal{L}\,\sqrt{|g|})}{\partial (\partial_{\mu}\psi^*)} = 0 \\
\frac{\partial (\mathcal{L}\,\sqrt{|g|})}{\partial A_{\mu}} - \frac{d}{d x^{\nu}}\frac{\partial (\mathcal{L}\,\sqrt{|g|})}{\partial (\partial_{\nu} A_{\mu})} = 0,
\end{align*}
we obtain the two coupled equations 
\begin{eqnarray}
\frac{\hbar^2}{2 m^*}(r^2s\ \partial_{rr}\psi + (2 r s + r^2s')\ \partial_r \psi)  \nonumber\\
+ \frac{\hbar^2}{2 m^*}(\nabla_{(x,y)} - i\frac{q}{\hbar} {\bf A})^2\psi \nonumber\\
 -\left(\frac{\alpha(T)}{r^2}+\frac{\chi}{L^2}+\beta\frac{|\psi|^2}{ r^k }\right)\ r^2\psi = 0
\label{GL1}
\end{eqnarray}
and
\begin{eqnarray}
{\bf J } &=& \frac{1}{r^2 \mu_0} \nabla_{(x,y)} \times \nabla_{(x,y)} \times {\bf A} - \frac{1}{\mu_0} (s'\partial_r {\bf A} + s\partial_{rr} {\bf A})  \nonumber\\
&=& \frac{q\hbar}{2m^*i} (\psi^*\nabla_{(x,y)}\psi -\psi \nabla_{(x,y)}\psi^* ) - \frac{q^2}{m^*}{\bf A}|\psi|^2.
\label{GL2}
\end{eqnarray}

Equation (\ref{GL1}) is a $2^{\rm nd}$ order differential equation for the scalar field, whereas Eq.(\ref{GL2}) is for a current density. Both equations incorporate a coupling of two parameters - the vector potential ${\bf A}$ and complex scalar field $\psi$. These equations describe the mechanics inside the bulk, where both gravitation and quantum mechanics co-exist \cite{Green}.

\subsection{Approximated Solutions}
Now we consider the complex scalar field $\psi(r,x,y)$, approximated by the following power series \cite{Gubser,Horowitz,Hartnoll,Garcia}
\begin{equation}
\psi(r,x,y) = \sum_{n=1}^{\infty} \frac{\psi_n(x,y)}{r^n} \approx \frac{\psi_1(x,y)}{r}+\frac{\psi_2(x,y)}{r^2}+\ldots,
\label{psi1}
\end{equation}
provided the holographic scale `$r$' is sufficiently large. 
Since the first two leading terms are linearly independent, one can choose any one of them to be an arbitrary solution of the gravitational system.
Similarly, the magnetic potential ${\bf A}$ can be chosen as
\begin{equation}
{\bf A}(r,x,y) \approx \left( 1 - b\exp{[-r/r_0]} \right){\bf A}_1(x,y),
\label{A1}
\end{equation}
which is the first order correction for the gauge field, where $b$ is a constant, and hence $\partial_r {\bf A} = b{\bf A}_1/(r_0\exp[r/r_0]$), $\partial_{rr} {\bf A} = -b{\bf A}_1/(r_0^2\exp[r/r_0])$.

Our approach focuses upon an extreme case where $r\rightarrow \infty$. This is where only quantum mechanics survives at the boundary of AdS space \cite{Maldacena1,Klebanov}. For a choice of $\psi = {\psi_1}/{r}$, it follows that $\partial_r \psi = - {\psi_1}/{r^2} $ and $\partial_{rr} \psi =  2{\psi_1}/{r^3}$.

\subsection{Choice of Scalar Solution ${\psi_1}/{r}$ (At the Infinite Boundary of AdS space: $r\rightarrow \infty$)}
For the case of `$r$' tending to a sufficiently large value $\Delta$ (where $\Delta \gg r_0$), we obtain $s(\Delta) = {\Delta^2}/{L^2}$ and $s'(\Delta) = 2{\Delta}/{L^2}$. The scalar-field equation (\ref{GL1}) then takes the following form
\begin{eqnarray}
\frac{\hbar^2}{2 m^*}(\nabla_{(x,y)} - i\frac{q}{\hbar}{\bf A}_1)^2\psi_1 = \nonumber\\
\left(\frac{\alpha(T)}{\Delta^2} + \frac{\chi}{L^2}+\frac{1}{L^2}\frac{\hbar^2}{m^*}+\beta\frac{|\psi_1|^2}{ \Delta^{k+2} }\right)\Delta^2\psi_1
\end{eqnarray}
For the choice of $\chi = -\hbar^2/m^*$ and the exponent $k=0$, the differential equation (\ref{GL1}) can reduce to  
\begin{equation}
 -\frac{\hbar^2}{2 m^*}\left(\nabla_{(x,y)} - i\frac{q}{\hbar}{\bf A}_1\right)^2\psi_1 + \alpha(T)\psi_1 + \beta |\psi_1|^2\psi_1  = 0,
 \label{BGL1}
\end{equation}
which is exactly the same as the $1^{\rm st}$ non-linear differential equation of Ginzburg-Landau theory \cite{Ginzburg}, where the coherence length is $\xi = \sqrt{{\hbar^2}/(2 m^* |\alpha(T)|)}$ and $\alpha(T)$, in this case, can be approximated as $\alpha(T) = \alpha_0 ({T-T_c})/{T_c}$ near the phase transition.

This arbitrary choice will automatically eliminate the term of $\Delta^2/L^2$, in which the superconducting system (as governed by quantum mechanics) survives in the absence of gravity. Our choice is similar to some proposals of negative potential $V=-2|\psi|^2/L^2$ which cancels the $\Delta^2/L^2$ term~\cite{Gubser,Horowitz,Hartnoll,Garcia}. We also preserve the phase transition property of the Ginzburg-Landau model, by introducing $\alpha(T)$ and $\beta$ within the potential $V(\psi)$.

Since the case of $r=\Delta$ is very large and close to the infinite boundary, applying the L'Hospital Rule, Eq. (\ref{GL2}) can reduce to 
\begin{eqnarray}
{\bf J } &=& \frac{1}{\mu_0}\nabla_{(x,y)} \times \nabla_{(x,y)} \times {\bf A}_1  \nonumber\\
&=& \frac{q\hbar}{2m^*i} (\psi_1^*\nabla_{(x,y)}\psi_1 -\psi_1 \nabla_{(x,y)}\psi_1^* ) \nonumber\\
&-& \frac{q^2}{m^*}{\bf A}_1|\psi_1|^2.
\label{BGL2}
\end{eqnarray}
Again, this is exactly the same as the $2^{\rm nd}$ differential equation of Ginzburg-Landau theory, describing the superconducting current \cite{Ginzburg}. We have now verified that the gravity model on the infinite boundary ($r\rightarrow \infty$) of AdS, can precisely emulate the Ginzburg-Landau theory in Euclidean space. This means that the superconducting system can be explained on this boundary.

\subsection{The Physical Meaning of $\psi$}
Finally, it is important to discuss about the physical meaning of $\psi$ both in the gravity and superconducting models. As is mentioned above, the scalar field $\psi$ in the bulk (described by the gravity model) now transits to the wave function $\psi = |\psi|\exp(i\theta)$ in the Ginzburg-Landau model at the AdS boundary. The bulk scalar field $\psi$ is analogue to the expectation value of source $<\mathcal{O}>$ in the quantum field theory. From the literature \cite{Gubser,Horowitz,Hartnoll,Garcia,Green,Ruth}, an arbitrary choice of $\psi$ in quantum theory could be either ${\psi_1}/{r}$ or ${\psi_2}/r^2$. In this study, the approximated solution is further limited to ${\psi_1}/{r}$ with the proposed potential $V(r,\psi)$, in a certain case ($\chi = -\hbar^2/m^{*}$ and $k=0$). This special choice of solution directly leads to the special coincidence of the gravity model in the AdS, to the superconducting theory at the AdS boundary. This result provides us with confidence to apply the mathematical techniques in quantum gravity to the superconductors (condensate matter physics).

\section{Conclusion and Remarks}
In brief, we have mathematically established the relationship between the (N+1) dimensional gravity model in the bulk AdS, to the N-dimensional Ginzburg-Landau system at the infinity Ads boundary ($r\rightarrow \infty$). It is found that the two coupled differential Ginzburg-Landau equations at the Ads boundary, can be derived from the equations of motion residing inside the bulk of AdS space. We restrict our efforts to the two-dimensional, asymptotically flat spatial domain of AdS space (as $r\rightarrow \infty$), and also propose the potential $V(r,\psi)$ for a special solution. The quantum gravity model is thought to expose many features that appear in the quantum critical electrons in the cuprate superconductors. As such, simplified models to analyse the nature of the condensed state using the AdS to Ginzburg-Landau formulation could lead to valuable new insights both in superconductivity, and black-hole physics. 









\input{jhepMay2016.bbl}
\end{document}

%% file: jhepMay2016.bbl
\providecommand{\href}[2]{#2}\begingroup\raggedright\endgroup